%% file: main.tex
\documentclass[11pt,a4paper]{article}
\usepackage[left=2.35cm,top=3cm,bottom=3cm,right=2.3cm]{geometry}
\usepackage{graphicx}
\usepackage{amsmath}
\usepackage{amssymb}
\usepackage{cite}
\usepackage{tikz}
\usepackage[colorlinks=true
,urlcolor=blue
,anchorcolor=blue
,citecolor=blue
,filecolor=blue
,linkcolor=blue
,menucolor=blue
,linktocpage=true
,pdfproducer=medialab
]{hyperref}
\input{orcidlink}

\def\cO{\mathcal{O}}
\newcommand\GeV{{\rm GeV}}
\newcommand\MeV{{\rm MeV}}
\newcommand\mcmule{{\sc McMule}}

\begin{document}

\begin{center}
{\Large\bf Low-energy {\boldmath $e^+\,e^-\to\gamma\,\gamma$} at NNLO in QED}
\\[2em]
{
\large Tim~Engel$^{a}$\,\orcidlink{0000-0003-2794-9032}\,, 
Marco~Rocco$^{b}$\,\orcidlink{0000-0002-2561-1209}\,, 
Adrian~Signer$^{c,d}$\,\orcidlink{0000-0001-8488-7400}\,, 
and Yannick~Ulrich$^{e}$\,\orcidlink{0000-0002-9947-3064}\,
}\\[0.2in]
{\sl ${}^a$ Kantonsschule Menzingen, 6313 Menzingen, Switzerland\\
\sl ${}^b$ Universit\`a degli Studi di Torino \& INFN, 10125 Torino, Italy \\
\sl ${}^c$ PSI Center for Neutron and Muon Sciences, 5232 Villigen PSI, Switzerland \\
\sl ${}^d$ Physik-Institut, Universit\"at Z\"urich, CH-8057 Z\"urich, Switzerland \\
\sl ${}^e$ University of Liverpool, Liverpool L69 3BX, U.K.
}
\setcounter{footnote}{0}
\end{center}

\vspace{0.2em}

\begin{center}
\begin{minipage}{0.9\textwidth}
{\small We present a fully differential computation of $e^+\,e^-\to\gamma\,\gamma$ at next-to-next-to-leading order in QED. The process has been implemented into \mcmule{}, completing its set of next-to-next-to-leading-order calculations for the most important $2\to{2}$ processes. The results allow for generic applications to electron-positron colliders with centre-of-mass energies up to a few GeV, particularly for luminosity measurements.}
\end{minipage}
\end{center}
\vspace{2em}

\newpage

\input{introduction}
\input{calculation}
\input{results}
\input{conclusion}

\subsection*{Acknowledgements}
It is a pleasure to thank Luca Naterop for his contributions at the initial stage of this project. 
We would like to thank Carlo Carloni Calame and Fulvio Piccinini for discussions regarding the comparison of our NNLO results to their NLO+PS results, and Graziano Venanzoni for explanations regarding the beam-energy spread at KLOE.
We acknowledge support by the Swiss National Science Foundation (SNSF) under grant 207386, and by the Italian Ministry of Universities and Research (MUR) through grants PRIN 2022BCXSW9 and FIS (CUP: D53C24005480001, FLAME).

\bibliographystyle{JHEP}
\bibliography{main}

\end{document}

%% file: orcidlink.tex
\usetikzlibrary{svg.path}

\definecolor{orcidlogocol}{HTML}{A6CE39}
\newcommand\orcidlink[1]{\href{https://orcid.org/#1}{%
\begin{tikzpicture}[yscale=-0.026,xscale=0.026,transform shape]
    \fill[orcidlogocol] svg{M256,128c0,70.7-57.3,128-128,128C57.3,256,0,198.7,0,128C0,57.3,57.3,0,128,0C198.7,0,256,57.3,256,128z};
    \fill[white] svg{M86.3,186.2H70.9V79.1h15.4v48.4V186.2z}
                 svg{M108.9,79.1h41.6c39.6,0,57,28.3,57,53.6c0,27.5-21.5,53.6-56.8,53.6h-41.8V79.1z M124.3,172.4h24.5c34.9,0,42.9-26.5,42.9-39.7c0-21.5-13.7-39.7-43.7-39.7h-23.7V172.4z}
                 svg{M88.7,56.8c0,5.5-4.5,10.1-10.1,10.1c-5.6,0-10.1-4.6-10.1-10.1c0-5.6,4.5-10.1,10.1-10.1C84.2,46.7,88.7,51.3,88.7,56.8z};
\end{tikzpicture}\!}%
}

%% file: introduction.tex
\section{Introduction} \label{sec:introduction}

The annihilation of an electron-positron pair into two photons, $e^+\,e^-\to\gamma\,\gamma$, for small centre-of-mass energy, is a classical QED process. It can be used for luminosity measurements~\cite{WorkingGrouponRadiativeCorrections:2010bjp} and, therefore, belongs to the program of several current and future low-energy experiments~\cite{Belle-II:2018jsg, BESIII:2017lkp, BESIII:2022ulv}. It also plays a role as a background process in the search for dark photons~\cite{Raggi:2014zpa}.  Given the relevance of low-energy processes at electron-positron colliders, there has been substantial effort and progress on the theoretical side~\cite{Aliberti:2024fpq}. In particular, the developments concern the inclusion of hadronic effects and corrections beyond next-to-leading order (NLO) in QED. 

The NLO QED corrections to $e^+\,e^-\to\gamma\,\gamma$ have been reported long ago~\cite{Andreassi:1962abc, Eidelman:1978rw, Berends:1980px} but a complete and corrected result was presented only more recently~\cite{Lee:2020zpo}. Direct comparisons with experiments however require more differential calculations~\cite{Arbuzov:1997pj}. The current state-of-the art for fully differential cross sections is NLO+PS, provided by BabaYaga~\cite{Balossini:2008xr}. It includes the full NLO corrections combined with a resummation of collinear emission through a parton shower.
While this approach does not include vacuum-polarisation (VP) contributions, their effect has been estimated in~\cite{CarloniCalame:2019dom} for higher energies. This work also includes electroweak corrections that are relevant for higher energies, but negligible for low-energy diphoton production.

In this article, we present a fully differential calculation of $e^+\,e^-\to\gamma\,\gamma$ at next-to-next-to-leading order (NNLO) in QED, including all photonic corrections, as well as fermion-loop corrections and dominant hadronic contributions. This calculation is complementary to the BabaYaga NLO+PS calculation. The NNLO results presented here include terms that are not included in~\cite{Balossini:2008xr}, namely NNLO terms beyond the parton-shower approach. On the other hand, the NLO+PS results include terms that are not present in our calculation, namely collinear enhanced terms beyond NNLO. A comparison between the two approaches offers a possibility for mutual cross checks and a solid assessment of the theoretical uncertainties due to missing higher-order effects.

In the context of processes involving photon pairs, it is worth mentioning available NLO electroweak calculations for $\gamma\gamma$ into light or heavy fermions~\cite{Denner:1995ar, Denner:1998tb}. On the other hand, diphoton production from a quark pair at NNLO in QCD has also received a lot of attention. Following two-loop results with massless internal quarks~\cite{Catani:2011qz, Catani:2018krb, Campbell:2016yrh, Schuermann:2022qdm}, calculations with massive internal quarks and phenomenological results have been presented recently~\cite{Becchetti:2023yat, Becchetti:2023wev, Becchetti:2025rrz, Ahmed:2025osb, Coro:2023mwl}. Even three-loop amplitudes with massless quarks have been considered~\cite{Caola:2020dfu}. There is, however, a crucial difference between QED and QCD calculations. In QED, the mass of the external fermions (i.e.~the electrons) must not be neglected, whereas in all QCD computations, the external quarks are treated as massless. The difference originates from the fact that QED is more exclusive with respect to collinear emission.

Our calculation has been done in the \mcmule{} framework~\cite{Banerjee:2020rww,Ulrich:2020frs}, using techniques developed in the context of Bhabha~\cite{Banerjee:2021mty}, M{\o}ller~\cite{Banerjee:2021qvi}, muon-electron~\cite{Broggio:2022htr}, and lepton-proton~\cite{Engel:2023arz} scattering. In particular, we use OpenLoops~\cite{Buccioni:2017yxi, Buccioni:2019sur} for the real-virtual corrections. The inclusion of external mass effects leads to technical complications in the two-loop amplitude. Since the corresponding massive amplitudes are not yet available, we use massified amplitudes, neglecting polynomially suppressed mass terms in the double-virtual contribution. The only additional approximation we make is to include only massless electrons for the light-by-light contributions. As we will show, these approximations are very well justified for the centre-of-mass energies $\sqrt{s}\sim \text{few}\,\GeV$ we are interested in. 

A more detailed account of the included contributions and their calculation is given in Section~\ref{sec:calculation}. Our results are shown in Section~\ref{sec:results}. First, we contrast our NNLO results for the total cross section with certain cuts to previous results in the literature. Next, we show differential results for experimental scenarios inspired by the KLOE and Belle II experiments. We stress that these are just illustrations. The \mcmule{} code can be used for generic low-energy scenarios with two detected photons. In Section~\ref{sec:conclusion} we present our conclusions.

%% file: calculation.tex
\section{Calculation} \label{sec:calculation}

\begin{figure}[h]
\centering
\includegraphics[width=0.95\textwidth]{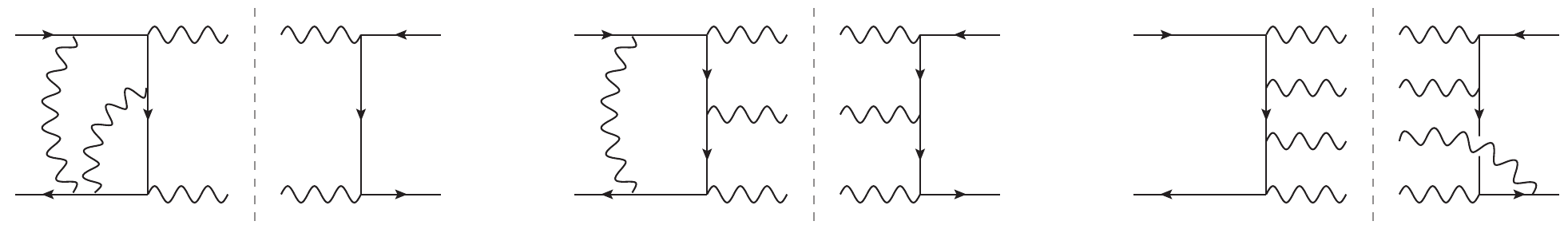}
\caption{Representative diagrams of the squared matrix element for the double-virtual (left), real-virtual (middle), and double-real (right) contributions to the photonic corrections $\sigma^{(2,\gamma)}$.
\label{fig:diagY}}
\end{figure}

\begin{figure}[h]
\centering
\includegraphics[width=0.95\textwidth]{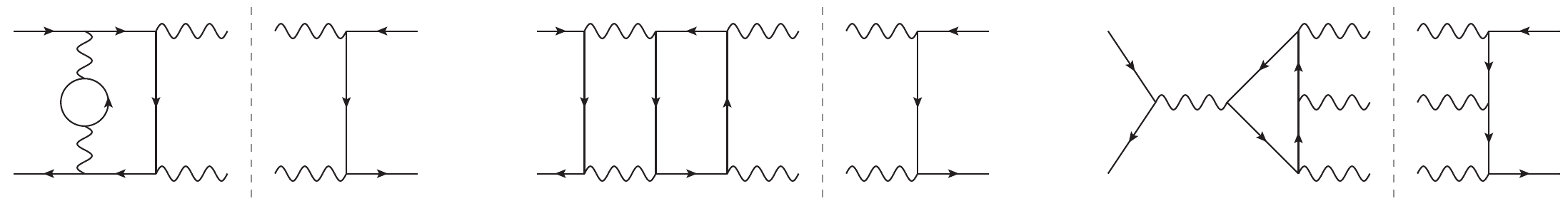}
\caption{Representative diagrams of the squared matrix element contributing to $\sigma^{(2,\text{VP})}$ (left), $\sigma^{(2,\text{LbL})}$ (middle), and $\sigma^{(2,\text{rLBL})}$ (right). 
\label{fig:diagVP}}
\end{figure}

For the calculation of $e^+\,e^-\to\gamma\,\gamma$ at NNLO we exploit the \mcmule{} framework~\cite{Banerjee:2020rww,Ulrich:2020frs} and follow closely the procedure used for other \mcmule{} NNLO computations~\cite{Banerjee:2021mty, Banerjee:2021qvi, Broggio:2022htr, Engel:2023arz}. We write the cross section at NNLO as
\begin{subequations}\label{eq:XSnnlo}
\begin{align}
\sigma_2&= \sigma_0+\sigma^{(1)} + \sigma^{(2)} = \sigma_1 + \sigma^{(2)}\\
\sigma^{(2)}&= \sigma^{(2,\gamma)} + \sigma^{(2,\text{VP}e)} +
   \sigma^{(2,\text{LbL}e)} + \sigma^{(2,\text{rLbL}e)}\\
\sigma^{(2+\text{VP})}&= \sigma^{(2)} +
   \sigma^{(2,\text{VP}\mu\tau)} +
   \sigma^{(2,\text{VPh})}
\end{align}
\end{subequations}   
The NLO corrections $\sigma^{(1)} = \sigma^{(1,\gamma)}$ are purely photonic, i.e.~there are no fermion loops. At NNLO, the corrections $\sigma^{(2)}$ are split into purely photonic contributions $\sigma^{(2,\gamma)}$, illustrated in Figure~\ref{fig:diagY}, and contributions with fermion loops, illustrated in Figure~\ref{fig:diagVP}. The latter can be split further into vacuum-polarisation (VP) contributions $\sigma^{(2,\text{VP}x)}$, and light-by-light (LbL) contributions $\sigma^{(2,\text{LbL}x)}$ and $\sigma^{(2,\text{rLbL}x)}$. Here, $x$ labels the particle in the loop and $\sigma^{(2)}$ corresponds to the NNLO corrections in QED with only electrons. Contributions from $x\neq{e}$ are added in $\sigma^{(2+\text{VP})}$. The fermion-loop contributions are dominated by electron loops $x=e$, but in principle we also have to take into account other leptons $x\in\{\mu,\tau\}$, as well as hadrons $x=h$ in the loop. As we will argue below, for the low-energy scenarios we are considering here, the light-by-light contributions are strongly suppressed and it is more than sufficient to consider only electrons in the loop.

For the photonic corrections we have the usual split into double virtual (including one-loop squared), real virtual, and double real, indicated in Figure~\ref{fig:diagY}. As discussed in more detail in~\cite{LN2021}, for the double virtual corrections we start from the amplitudes with a massless fermion~\cite{Anastasiou:2002zn} and use massification~\cite{Penin:2005eh, Becher:2007cu, Mitov:2006xs, Engel:2018fsb}. This procedure neglects terms polynomially suppressed by the electron mass $m_e$. Since all four external states are well separated and we consider centre-of-mass energies $\sqrt{s} \gg m_e$, the corresponding error is completely negligible. The real-virtual amplitudes are evaluated with full mass dependence using OpenLoops~\cite{Buccioni:2017yxi, Buccioni:2019sur}. For the double-real contributions we use an in-house evaluation of the matrix element that was checked against OpenLoops. The loop amplitudes are renormalised in the on-shell scheme and the infrared singularities are treated with the FKS$^\ell$ subtraction scheme~\cite{Engel:2019nfw}.

While there are different techniques to compute the VP contributions $\sigma^{(2,\text{VP}x)}$, we found it most convenient to use a dispersive approach (for a brief overview see~\cite{Aliberti:2024fpq,Fang:2025mhn}). This allows for an efficient inclusion of leptons $x\in\{e,\mu,\tau\}$ as well as hadrons $x=h$ in the loop. The latter contribution cannot be obtained using perturbation theory and we use the {\tt alphaQED23} package~\cite{Jegerlehner:2006ju, Jegerlehner:hvp19}. The two-loop VP part is separately infrared finite. 

This leaves us with the light-by-light contributions that can be further split into two separately infrared-finite parts, as shown in Figure~\ref{fig:diagVP}. The first, $\sigma^{(2,\text{LbL})}$, consists of double-box diagrams, with one box containing the fermion (or hadron) loop. The second, $\sigma^{(2,\text{rLbL})}$, can be considered as a real correction, as it has three photons in the final state. However, the limit of one of the photons becoming soft does not lead to a singularity. Both parts are included in our results with a massless electron in the loop. The computation with a massive lepton loop is significantly more involved. Given the minimal numerical impact of $x\in\{\mu, \tau\}$ for the light-by-light contributions, we drop these terms. Similarly, we also do not include hadronic light-by-light contributions. Their evaluation would be even more elaborate and require techniques similar to those used for the corresponding contributions to the anomalous magnetic moment of the muon~\cite{Aliberti:2025beg}. 

%% file: results.tex
\section{Results} \label{sec:results}

All results presented in the following are publicly available in the \mcmule{} user library~\cite{McMule:data}
\begin{quote}
  \url{https://mule-tools.gitlab.io/user-library/diphoton/nnlo-studies}
\end{quote}
The production runs employed version {\tt v0.7.0} of the \mcmule{} public release on Intel Xeon Gold 6140 CPUs. Timing details are specified for each configuration in the relevant sections. More details, e.g.~the number of evaluation points, can be found in the repository of the \mcmule{} User Library
\begin{quote}
  \url{https://gitlab.com/mule-tools/user-library/-/tree/master/diphoton/}
\end{quote}
The major time bottleneck is due to NNLO photonic corrections, requiring 80--90\% of the total computing time.

\subsection{Total cross sections} \label{sec:res-totXS}

We have compared the NLO total cross section without any cuts to the analytic results given in~\cite{Lee:2020zpo}. For very low energies $\sqrt{s} \lesssim 10\,\MeV$ it is straightforward to obtain Monte Carlo errors of $10^{-6}$, whereas for energies approaching $\sqrt{s}\simeq 10\,\GeV$ the Monte Carlo errors increase to $10^{-4}$. In all cases we have found agreement within these errors. In order to obtain the accuracy mentioned above, we ran \mcmule{} for 0.002 CPU-years for energies $\sqrt{s} \lesssim 100\,\MeV$, 0.3 CPU-years for energies $\sqrt{s} \gtrsim 1\,\GeV$.

We start the presentation of our NNLO results with the total cross section
for three values of $\sqrt{s}\in\{1,3,10\}\,\GeV$. In addition we
apply the following cuts
\begin{subequations}\label{eq:cutTotXS}
\begin{align}
    &\theta_\gamma^{\text{min}}= 45^\circ ,&    &\theta_\gamma^{\text{max}}= 135^\circ ,&   \\
    &E_\gamma^{\text{min}}= 0.3\,\sqrt{s} ,&    &\xi^{\text{max}}= 10^\circ .&
\end{align}
\end{subequations}
Thus, we request at least two photons with energy larger than
$E_\gamma^{\text{min}}$ in an angular range determined by
$\theta_\gamma^{\text{min}}$ and $\theta_\gamma^{\text{max}}$. The
acollinearity $\xi\equiv |\theta_+ + \theta_- - 180^\circ|$ of the two
most energetic photons satisfying these cuts must be smaller than
$\xi^{\text{max}}$. These cuts correspond precisely to cuts used
in~\cite{Balossini:2008xr}, with the results reported
in~\cite{WorkingGrouponRadiativeCorrections:2010bjp}.

Our results are summarised in Table~\ref{tab:totXS} and compared
to~\cite{Balossini:2008xr}. Independently of the centre-of-mass energy, each configuration required a running time of 0.06 CPU-years. The NLO results agree within 0.01\% and the corrections are between $6-8$\%. Beyond NLO, there are physical
differences between \cite{Balossini:2008xr} and our results. We have
included the full $\cO(\alpha^2)$ NNLO corrections, whereas
in~\cite{Balossini:2008xr} photonic corrections beyond NLO have been
implemented with a parton-shower approach. Thus, VP, LbL, and some
non-logarithmic $\cO(\alpha^2)$ corrections are missing in
\cite{Balossini:2008xr}. On the other hand, they resum the
leading-logarithmic corrections from multiple photon emission and,
hence, include dominant terms of photonic corrections also beyond
NNLO. Rather than comparing the best predictions it is thus more
informative to compare the exact photonic NNLO corrections
$\sigma^{(2,\gamma)}$ with the parton-shower induced contributions
beyond NLO. In the notation of
\cite{Balossini:2008xr}, where $\sigma^\text{NLO}_\alpha$ is used for $\sigma_1$, they are given by
$\sigma_\text{exp}-\sigma_1$. As can be seen from
Table~\ref{tab:totXS}, the two approaches agree within $10-20$\% and
give beyond-NLO photonic corrections of the order of $0.2-0.5$\%.
Taking into account the $\cO(\alpha^2)$ suppression, this results in
differences of the order of 0.05\%.

In \mcmule{} also further NNLO corrections are included. In
particular, there are VP corrections due to an electron loop,
$\sigma^{(2,\text{VP}e)}$. These corrections are negative and about
20\% of $\sigma^{(2,\gamma)}$, becoming slightly more important for
increasing $\sqrt{s}$. Including $\sigma^{(2,\text{VP}e)}$ in the
\mcmule{} result leads to near-perfect agreement with the
parton-shower result, but this agreement is accidental. The VP
corrections with heavier particles in the loop,
$\sigma^{(2,\text{VP}\mu\tau)}$ and $\sigma^{(2,\text{VP}h)}$, are
strongly suppressed for the small centre-of-mass energies considered
here. Even for $\sqrt{s}=10\,\GeV$ they are about a factor 100 smaller
than $\sigma^{(2,\text{VP}e)}$. Thus, they can be safely neglected. In
fact, they are smaller than the numerical differences of the NLO
results. This also holds for light-by-light contributions with an
electron in the loop, $\sigma^{(2,\text{LbL}e)}$ and
$\sigma^{(2,\text{rLbL}e)}$. These results fully justify the neglect
of the corresponding contributions with heavier leptons or hadrons in
the loop.

\begin{table}[th]
  \renewcommand*{\arraystretch}{1.2}
  \begin{center}
    \begin{tabular}{l|rr|rr|rr}
      $\sqrt{s}$ & \multicolumn{2}{c|}{1\,GeV}
      & \multicolumn{2}{c|}{3\,GeV} & \multicolumn{2}{c}{10\,GeV} \\
      & {\tiny \mcmule{} [nb]}\phantom{aaa}
      & {\tiny \cite{Balossini:2008xr}  [nb]}
      & {\tiny \mcmule{} [nb]}\phantom{a}
      & {\tiny \cite{Balossini:2008xr}  [nb]}
      & {\tiny \mcmule{} [nb]}\phantom{a}
      &{\tiny \cite{Balossini:2008xr}  [nb]} \\
      \hline
      $\sigma_0$  &
      137.531\phantom{000} & 137.53 &
      15.2812\phantom{0}   & 15.281 &
      1.37531\phantom{0}   & 1.3753 \\
      $\sigma_1$ &
      129.444\phantom{000} & 129.45 &
      14.2099\phantom{0}   & 14.211 &
      1.26185\phantom{0}   & 1.2620 \\
      $\sigma_2$ &
      129.760\phantom{000} &  &
      14.2570\phantom{0}   &  &
      1.26738\phantom{0}   &  \\
      $\sigma_{\text{exp}}$ &
      & 129.77 &
      & 14.263  &
      & 1.2685  \\
      \hline   
      $\sigma^{(2,\gamma)}$ &
      0.383\phantom{000} &  \phantom{{\huge A}} &
      0.0598\phantom{0}   &  &
      0.00738\phantom{0}   &  \\  
      $\sigma_\text{exp}-\sigma_1$ & 
      & 0.32 &
      & 0.053 &
      & 0.0065 \\  
      \hline
   $\sigma^{(2,\text{VP}e)}$ &
     $-$0.069\phantom{000} & \phantom{{\huge A}} &
     $-$0.0128\phantom{0} & &
     $-$0.00186\phantom{0} & \\
   $\sigma^{(2,\text{LbL}e)}$ &
     $-$0.0014\phantom{00} & &
     $-$0.00016 & &
     $-$0.000014 & \\
   $\sigma^{(2,\text{rLbL}e)}$ &
     0.0030\phantom{00} & &
     0.00033 & &
     0.000030 & \\
   $\sigma^{(2,\text{VP}\mu\tau)}$ &
     $-$0.000077 & &
     $-$0.00018 & &
     $-$0.000080 & \\
   $\sigma^{(2,\text{VPh})}$ &
     0.000090 & &
     $-$0.00010 & &
     $-$0.000097 &      
    \end{tabular}
    \caption{Different contributions to the NNLO cross section computed
      by \mcmule{} and compared to results in
      \cite{Balossini:2008xr}. Regarding the latter,
      $\sigma_\text{exp}$ corresponds to matching a complete NLO
      result with a parton-shower resummation.  \label{tab:totXS}}
  \end{center}
\end{table}

To summarise the situation for the total cross section we confirm the
estimate of the theoretical accuracy of 0.1\% given
in~\cite{Balossini:2008xr}.  We continue the presentation of our
results with differential distributions in two different scenarios,
inspired by the KLOE and BELLE II experiments.

\subsection{Differential cross sections for a KLOE-like scenario}
\label{sec:res-KLOE}

The centre-of-mass energy is set to $\sqrt{s} = 1.0\,\GeV$ and we
require at least two photons with
\begin{align}
  \label{kloeycut}
&E_\gamma \ge 300\,\MeV &  &\mbox{and}&
&45^\circ \le \theta_\gamma \le 135^\circ\, .&
\end{align}
In addition we apply the cut on the acollinearity, requiring
\begin{align}
  \label{kloegcut}
  \xi &\equiv |\theta_l+\theta_s-\pi| \le 10^\circ\, ,
\end{align}
where $\theta_l$ and $\theta_s$ are the angles of the two most
energetic photons satisfying \eqref{kloeycut}. We apply a beam spread
for both incoming particles, with a Gaussian distribution of FWHM
0.12\,\MeV~\cite{KLOE:2004uzx}. A beam spread is able to cure largely
fluctuating distributions caused by soft enhancements when the
energies of tree-level photons are equal. Further, the sorting
algorithm for photons in the final state requires a percent-level
energy resolution, otherwise the event is randomly
ordered. The results obtained for the KLOE-like scenario required a running time of 0.07 CPU-years.

\begin{figure}[h]
\centering
\includegraphics[width=0.9\textwidth]{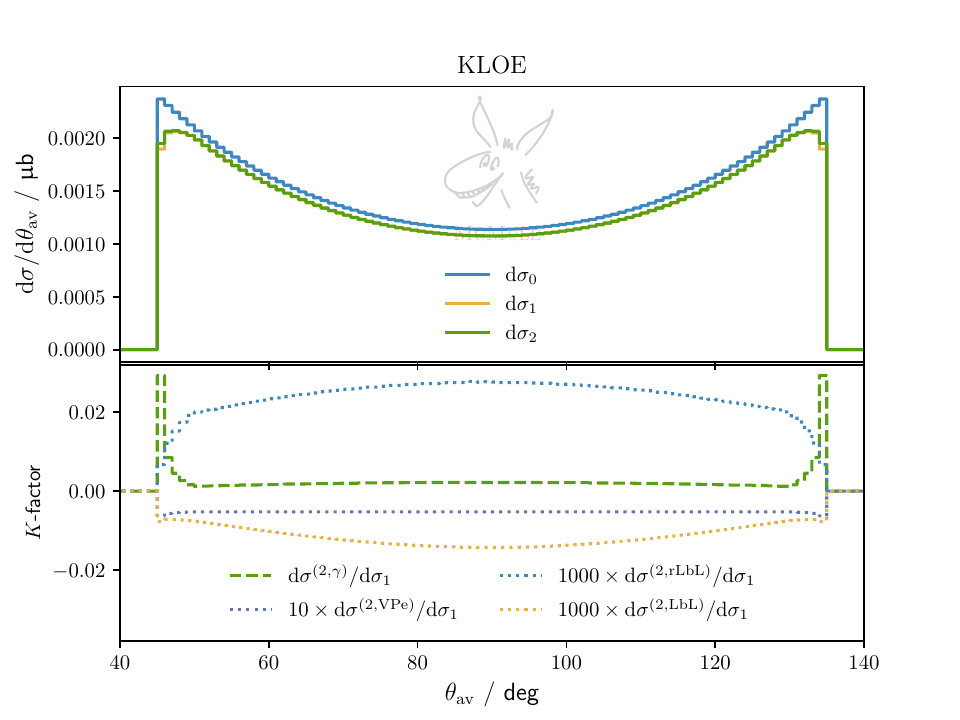}
\caption{Differential distribution with respect to $\theta_{\rm av}$
  for the KLOE-like scenario. The blue, orange, and green curves in
  the upper panel correspond to LO-, NLO-, and NNLO-precise
  distributions, respectively. The size of NNLO photonic and
  non-photonic corrections are presented in the lower panel with
  respect to the NLO distribution, with varying enhancing factors for
  the latter. }
\label{fig:kloe-thav}
\end{figure}

The differential distribution with respect to
\begin{equation}
  \label{eq:thav}
  \theta_{\rm av} \equiv \frac{1}{2} (\theta_l-\theta_s+\pi)
\end{equation}
is shown in Figure~\ref{fig:kloe-thav}. The symmetrical shape is due to
the symmetrical beam energies. NNLO corrections amount to a few
permille, with strong enhancements at the edges of the distribution,
up to 2\%. The lower panel confirms the negligibility of non-photonic
corrections compared to photonic corrections, particularly for this
low-energy KLOE-like scenario. The by-far biggest contribution among
non-photonic corrections is due to $\sigma^{(2,\text{VP}e)}$,
i.e.~diagrams with an electronic vacuum polarisation insertion. While
these are computed with no approximations, $\sigma^{(2,\text{LbL}e)}$
and $\sigma^{(2,\text{rLbL}e)}$ are computed for electronic insertions
only, further with a massless electron. The size of the latter
corrections justifies the approximation, hence the allusion in
Section~\ref{sec:calculation}.

\subsection{Differential cross sections for a Belle-like scenario}
\label{sec:res-Belle}

The process is $e^-(7\,\GeV)\, e^+(4\,\GeV)\to \gamma\,\gamma$ with at
least two photons satisfying
\begin{align}
  \label{belleycut}
    &E_\gamma \ge 1\,\GeV& &\mbox{and}&  &15^\circ \le \theta_\gamma \le 165^\circ\, .&
\end{align}
Even though we use a symmetric angular cut, the parameters of the
photons are understood to be in the LAB frame. Hence, photons tend to be
emitted forward, i.e.~$\theta_\gamma > 90^\circ$. The centre-of-mass
energy is $\sqrt{s}=10.583\,\GeV$. No beam spread is applied in this
case, as the beam energies are asymmetrical by construction. The results obtained for the Belle-like scenario required a running time of 0.7 CPU-years.

\begin{figure}[h]
\centering
\includegraphics[width=0.9\textwidth]{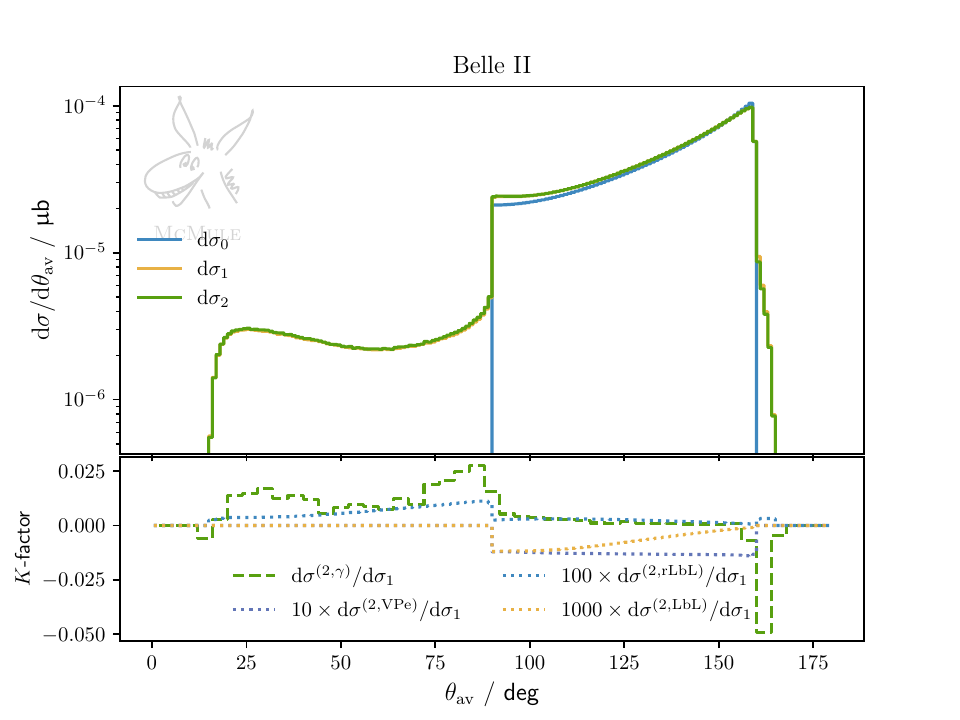}
\caption{Differential distribution with respect to $\theta_{\rm av}$
  for the Belle-like scenario. The blue, orange, and green curves in
  the upper panel correspond to LO-, NLO-, and NNLO-precise
  distributions, respectively. The size of NNLO photonic and
  non-photonic corrections are presented in the lower panel with
  respect to the NLO distribution, with varying enhancing factors for
  the latter. }
\label{fig:belle-thav}
\end{figure}

The differential distribution with respect to $\theta_{\rm av}$ is
presented for this scenario as well, and is shown in
Figure~\ref{fig:belle-thav}. NNLO corrections amount to a few permille,
with enhancements below $\sim 85^\circ$ and above $\sim 160^\circ$,
where the LO contribution is zero and the NNLO calculation effectively
provides an NLO result. The size of non-photonic corrections is larger
than in the KLOE-like scenario, due to the higher centre-of-mass
energy.  The biggest contribution among non-photonic corrections is
due to $\sigma^{(2,\text{VP}e)}$, i.e.~diagrams with an electronic
vacuum polarisation insertion, whose size grows at higher
energies. This term is of the order of 0.1\%, so should be included
according to the accuracy goal. Instead, $\sigma^{(2,\text{LbL}e)}$
and $\sigma^{(2,\text{rLbL}e)}$, as well as contributions with heavier
particles in the loop, require large enhancing factors in order to
become visible on the plot.

%% file: conclusion.tex
\section{Conclusion} \label{sec:conclusion}

We have presented a complete NNLO QED calculation for photon pair production at low-energy electron-positron colliders. The results have been implemented into \mcmule{} and as an illustration we have shown distributions for scenarios inspired by the KLOE and Belle experiments. However, we stress that the code allows the computation of any differential cross section with arbitrary (infrared-safe) cuts. Since the computation is pure QED, the centre-of-mass energy has to be sufficiently small to justify the neglect of electroweak effects. 

This calculation is based on techniques developed in the context of previous \mcmule{} NNLO computations.  It completes the set of fully differential NNLO calculations for the most important $2\to{2}$ processes in \mcmule. From the current results it is in principle also possible to obtain Compton scattering at NNLO through crossing.

The comparison of the photonic NNLO calculation of $e^+\,e^-\to\gamma\,\gamma$ to previous NLO+PS results confirms earlier error estimates of the order of 0.1\%. The fermionic corrections are of a similar size. This is a clear indication that terms beyond NNLO as well as the terms beyond NLO+PS have an impact at most at the permille level. This comparison is indeed one of the most reliable ways to estimate the effect of missing higher-order corrections. It will be very interesting to perform a similar comparison for $2\to{3}$ processes. With the available techniques, there are no decisive obstacles towards this goal.